\documentclass[submission, Phys]{SciPost}
\usepackage{authblk}
\usepackage{tabularx}

\usepackage{url}
\usepackage{longtable}
\usepackage{multirow}
\usepackage[font=small,skip=0pt]{caption}
\usepackage[labelformat=simple]{subcaption}

\usepackage[nodayofweek]{datetime}
\usepackage{enumerate}
\usepackage{xspace}
\usepackage{xcolor}
\usepackage{graphicx}
\usepackage[export]{adjustbox}
\usepackage{algorithm}
\usepackage{algpseudocode}
\usepackage{bookmark}

\usepackage{float}
\usepackage[capitalise]{cleveref}
\usepackage{algorithm}
\usepackage{algpseudocode}
\usepackage{placeins}
\usepackage{booktabs}

\newcolumntype{Y}{>{\centering\arraybackslash}X}
\newcolumntype{R}{>{\raggedleft\arraybackslash}X}

\newcommand{\SBone}{SB1\xspace}
\newcommand{\SBtwo}{SB2\xspace}

\newcommand{\CURTAINs}{\mbox{\textsc{Curtain}s}\xspace}
\newcommand{\FfF}{\mbox{\textsc{Curtain}sF4F}\xspace}
\newcommand{\CATHODE}{\mbox{\textsc{Cathode}}\xspace}
\newcommand{\FETA}{\textsc{Feta}\xspace}
\newcommand{\CWoLa}{\mbox{\textsc{CWoLa}}\xspace}
\newcommand{\mjj}{\ensuremath{m_{JJ}}\xspace}
\newcommand{\nvidia}{NVIDIA\textsuperscript{\textregistered}\xspace}

\hypersetup{
    colorlinks,
    linkcolor={red!50!black},
    citecolor={blue!50!black},
    urlcolor={blue!80!black}
}

\usepackage[bitstream-charter]{mathdesign}
\DeclareSymbolFont{usualmathcal}{OMS}{cmsy}{m}{n}
\DeclareSymbolFontAlphabet{\mathcal}{usualmathcal}

\begin{document}

\begin{center}{\Large \textbf{
            CURTAINs Flows For Flows:\\ Constructing Unobserved Regions with Maximum Likelihood Estimation
        }}\end{center}

% TODO: write the author list here. Use first name (+ other initials) + surname format.
% Separate subsequent authors by a comma, omit comma and use "and" for the last author.
% Mark the corresponding author with a superscript star.
\begin{center}
    Debajyoti Sengupta,
    Samuel Klein,
    John Andrew Raine$^\star$, and
    Tobias Golling
\end{center}

\begin{center}
    University of Geneva\\
    {\small \sf debajyoti.sengupta@unige.ch} {\small \sf samuel.klein@unige.ch} {\small \sf john.raine@unige.ch} {\small \sf tobias.golling@unige.ch}
\end{center}

\begin{center}
    \today
\end{center}

% For convenience during refereeing (optional),
% you can turn on line numbers by uncommenting the next line:
%\linenumbers

\section*{Abstract}
 {\bf
  % TODO: write your abstract here.
  Model independent techniques for constructing background data templates using generative models have shown great promise for use in searches for new physics processes at the LHC.
  We introduce a major improvement to the \CURTAINs method by training the conditional normalizing flow between two side-band regions using maximum likelihood estimation instead of an optimal transport loss.
  The new training objective improves the robustness and fidelity of the transformed data and is much faster and easier to train.

  We compare the performance against the previous approach and the current state of the art using the LHC Olympics anomaly detection dataset, where we see a significant improvement in sensitivity over the original \CURTAINs method.
  Furthermore, \FfF requires substantially less computational resources to cover a large number of signal regions than other fully data driven approaches.
  When using an efficient configuration, an order of magnitude more models can be trained in the same time required for ten signal regions, without a significant drop in performance.
 }

\vspace{10pt}
\noindent\rule{\textwidth}{1pt}
\tableofcontents\thispagestyle{fancy}
\noindent\rule{\textwidth}{1pt}
\vspace{10pt}

\section{Introduction}
The search for new physics phenomena is one of the cornerstones of the physics programme at the Large Hadron Collider~(LHC).
The unparalleled energy and intensity frontier provided by the LHC provides a huge range of phase space where new signatures may be observed.
The ATLAS~\cite{ATLAS:2008xda} and CMS~\cite{CMS:2008xjf} collaborations at the LHC perform a wide array of searches for new particles beyond the standard model (BSM) of particle physics.
Many of these searches target specific models,
however, due to the vast possibilities of models and particles, dedicated searches cannot be performed for all possible scenarios.

Model independent searches aim to provide a broad sensitivity to a wide range of potential BSM scenarios without targetting specific processes.
A key technique used in many searches is the bump hunt.
Under the assumption that a new BSM particle is localised to a certain mass value, a bump hunt scans over an invariant mass distribution looking for excesses on top of a smooth background.
Bump hunts were crucial in the observation of the Higgs boson by the ATLAS and CMS collaborations~\cite{ATLAS:2012yve,CMS:2012qbp}.
However, despite the success at finding the Higgs boson, there is little evidence for any BSM particles at either experiment~\cite{ATLAS:2021ilc,ATLAS:2021zqc,ATLAS:2021yfa,CMS:summary1,CMS:summary2,CMS:summary3}.
With advances in machine learning~(ML) many new model independent methods have been proposed to enhance the sensitivity to BSM physics~\cite{Aguilar-Saavedra:2017rzt,Hajer:2018kqm,Heimel:2018mkt,Farina:2018fyg,Cerri_2019,Roy:2019jae,Blance_2019,Jawahar:2021vyu,D_Agnolo_2019,De_Simone_2019,Letizia:2022xbe,DAgnolo:2019vbw,Kasieczka:2021xcg,Aarrestad:2021oeb,Dillon:2023zac,Chen:2022suv,Kasieczka:2022naq,Dillon:2022tmm,Finke:2022lsu} including approaches which aim to improve the sensitivity of the bump hunt itself~\cite{cwolabump,anode,cathode,Andreassen:2020nkr,Benkendorfer:2020gek,Raine:2022hht,Hallin:2022eoq,feta}.

In this work we improve upon the \CURTAINs approach~\cite{Raine:2022hht} by replacing the optimal transport loss used to train a flow between two complex distributions with maximum likelihood estimation.
In order to evaluate the likelihood of the complex distributions on either side of the normalizing flow, we use the \emph{Flows for Flows} technique introduced in Ref.~\cite{flows4flows} and applied to physics processes in Ref.~\cite{Mastandrea:2022vas}.
This new configuration is called \FfF.

We apply \FfF to the LHC Olympics (LHCO) R\&D dataset~\cite{lhco4536377}, a community challenge dataset for developing and comparing anomaly detection techniques in high energy physics~\cite{Kasieczka:2021xcg}.
We compare it to the previous iteration of \CURTAINs, as well as to a current state of the art data driven approach \CATHODE~\cite{cathode}.
We evaluate the performance both in terms of improved signal sensitivity, but also in the required computational time to train the background models for a number of signal regions.

\section{Dataset}
We evaluate the performance of \FfF using the LHC Olympics R\&D dataset.

The LHCO R\&D dataset~\cite{lhco4536377} comprises background data produced through QCD dijet production,
with signal events arising from the all-hadronic decay of a massive particle to two other massive particles
$W^\prime\rightarrow X(\rightarrow q\bar{q}) Y(\rightarrow q\bar{q})$, each
with masses $m_{W^\prime} = 3.5$~TeV, $m_{X} = 500$~GeV, and $m_{Y} = 100$~GeV.
Both processes are simulated with \texttt{Pythia}~8.219~\cite{Sjostrand:2007gs} and interfaced to \texttt{Delphes}~3.4.1~\cite{deFavereau:2013fsa} for detector simulation.
Jets are reconstructed using the anti-$k_\mathrm{T}$ clustering algorithm~\cite{Cacciari:2008gp} with a radius parameter $R=1.0$, using the \texttt{FastJet}~\cite{Cacciari:2011ma} package.
In total there are 1~million QCD dijet events and 100\, 000 signal events.

Events are required to have exactly two large radius jets, with at least one passing a transverse momentum requirement $p_\mathrm{T}^{J} > 1.2$~TeV.
Jets are ordered by decreasing mass.
In order to remove the turn on in the \mjj distribution arising from the jet selections, we only consider events with $\mjj > 2.8$~TeV.
To construct the training datasets we use varying amounts of signal events mixed in with the QCD dijet data.

To study the performance of our method in enhancing the sensitivity in a bump hunt, we use the input features proposed in Refs.~\cite{cwolabump,anode,cathode,Raine:2022hht}.
These are
\begin{equation*}
    m_{JJ}, m_{J_1}, \Delta m_{J} = m_{J_1} - m_{J_2}, \tau_{21}^{J_1}, \tau_{21}^{J_2}, \mathrm{and} \, \Delta R_{JJ},
\end{equation*}
where $\tau_{21}$ is the N-subjettinness~\cite{nsubjettiness} ratio of a large radius jet,
and $\Delta R_{JJ}$ is the angular separation of the two jets in the detector $\eta-\phi$ space.

\section{Method}
\FfF follows the same motivation and approach as the original \CURTAINs method presented in Ref.~\cite{Raine:2022hht}.
In bump hunt searches, data are categorised into non overlapping signal (SR) and side-band (SB) regions on a resonant distribution (\mjj).
In \CURTAINs, a conditional Invertible Neural Network (cINN) is trained to learn the mapping from data drawn from one set of \mjj values to a target set of values.
% The network is trained to transport data from one \mjj value to their equivalent values at another point on \mjj.
The cINN is trained using the SB regions and applied to transport data from the SB to the SR.

However, the approach improves upon \CURTAINs by using a maximum likelihood loss on the transported data instead of an optimal transport loss between the batch of data and a batch of target data.

\subsection{Flows for Flows architecture}

% \FfF is based on the architecture proposed in Ref.~\cite{flows4flows}.
% A normalizing flow trained with maximum likelihood estimation requires an invertible neural network and a base distribution with.
A normalizing flow trained with maximum likelihood estimation requires an invertible neural network and a base distribution with an evaluable density.
The standard choice for the base distribution is a standard normal distribution.
The loss function for training the normalizing flow $f_\phi(z)$ from some distribution $x\sim X$ to the base distribution $z\sim p_{prior}$ is given from the change of variables formula
\begin{equation*}
    \log p_{\theta, \phi} (x) = \log p_\theta (f_\phi^{-1}(x)) - \log \left| \det (J_{f_\phi^{-1}(x)}) \right|,
\end{equation*}
where $J$ is the Jacobian of $f_\phi$.
In the conditional case this extends to
\begin{equation}
    \label{eq:mle_loss}
    \log p_{\theta, \phi} (x | c) = \log p_\theta \left(f_{\phi(c)}^{-1}(x | c)\right) - \log \left| \det (J_{f_{\phi(c)}^{-1}(x | c)}) \right|,
\end{equation}
where $c$ are the conditional properties.

In \cref{eq:mle_loss}, the base density term $\log p_\theta \left(f_{\phi(c)}^{-1}(x | c)\right)$ introduces a problem for training \CURTAINs with maximum likelihood estimation.
As the network is trained between two regions sampled from some non-analytically defined distribution, the probability of the transformed data is unknown.
As a result, an optimal transport loss was used instead.

\begin{figure}[hbpt]
    \centering
    \includegraphics{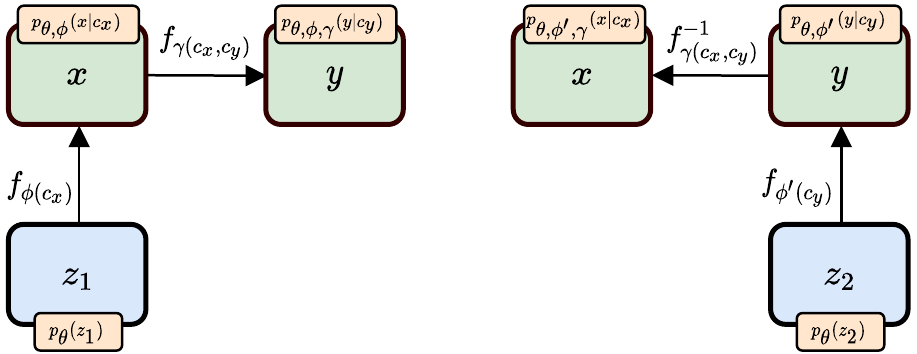}
    \caption{The Flows for Flows architecture for a conditional model~\cite{flows4flows}.
        Data $x$ ($y$) are drawn from the initial distribution with conditional values $c_x$ ($c_y$) and transformed to new values $c_y$ ($c_x)$ in a cINN $f_\gamma(c_x,c_y)$ conditioned on $c_x$ and $c_y$.
        The probability of the transformed data points are evaluated using a second normalizing flow for the base distribution $f_{\phi^\prime(c_y)}$ ($f_{\phi(c_x)}$).
        In the case where $x$ and $y$ are drawn from the same underlying distribution $p(x, c)$, the same base distribution $f_\phi$ can be used.
    }
    \label{fig:f4f}
\end{figure}

However, the base density of a normalizing flow can be chosen as any distribution for which the the density is known.
Therefore, we can train an additional normalizing flow to learn the conditional density $p_\theta \bigl(f_{\phi(c)}^{-1}(x|c)\bigr)$ of the target data distribution.
This second normalizing flow, the base distribution, is trained in advance and is used to define the loss in \cref{eq:mle_loss} when training the normalizing flow on arbitrary target distributions.
In \FfF the conditional properties of the top flow are a function of the input ($x$) and target ($y$) conditional properties $c_x$ and $c_y$.
For the base distribution only a single conditional property $c$ is needed.
% In \FfF the conditional properties are a function of the input and target \mjj values in ascending order, $m_{JJ}^{low}$ and $m_{JJ}^{high}$.
% This could be the concatenation of both, or the difference.
% For the base distribution, however, only the \mjj of the data is used.
The correspondence between the top normalizing flow and the base distributions in Flows for Flows is shown in \cref{fig:f4f}.

\subsection{Training \FfF}

As with the original training method, \FfF can be trained in both directions.
The forward pass transforms data from low to higher target values of \mjj, whereas the inverse pass transforms data from high to lower target values.
When training between two distinct arbitrary distributions in both directions, a base distribution is required for each distribution.

In principle, \FfF could be trained between data drawn from the low \mjj SB (\SBone) to the high \mjj SB (\SBtwo), as performed with \CURTAINs.
However, as data no longer need to be compared to a target batch, it is possible to train with both SBs combined in a simplified training.

Data are drawn from both SBs and target \mjj values ($m_\textit{target}$) are randomly assigned to each data point using all \mjj values in the batch.
Data are passed through the network in a forward or inverse pass depending on whether $m_\textit{target}$ is larger or smaller than their initial \mjj ($m_\textit{input}$).
The network is conditioned on a function of $m_\textit{input}$ and $m_\textit{target}$ with the two values ordered in ascending order ($f(m_{jj}^{low},m_{jj}^{high})$).
This function could be, for example, the concatenation of or difference between $m_{jj}^{low}$ and $m_{jj}^{high}$.

\begin{figure}[hbpt]
    \centering
    \includegraphics[width=0.5\textwidth]{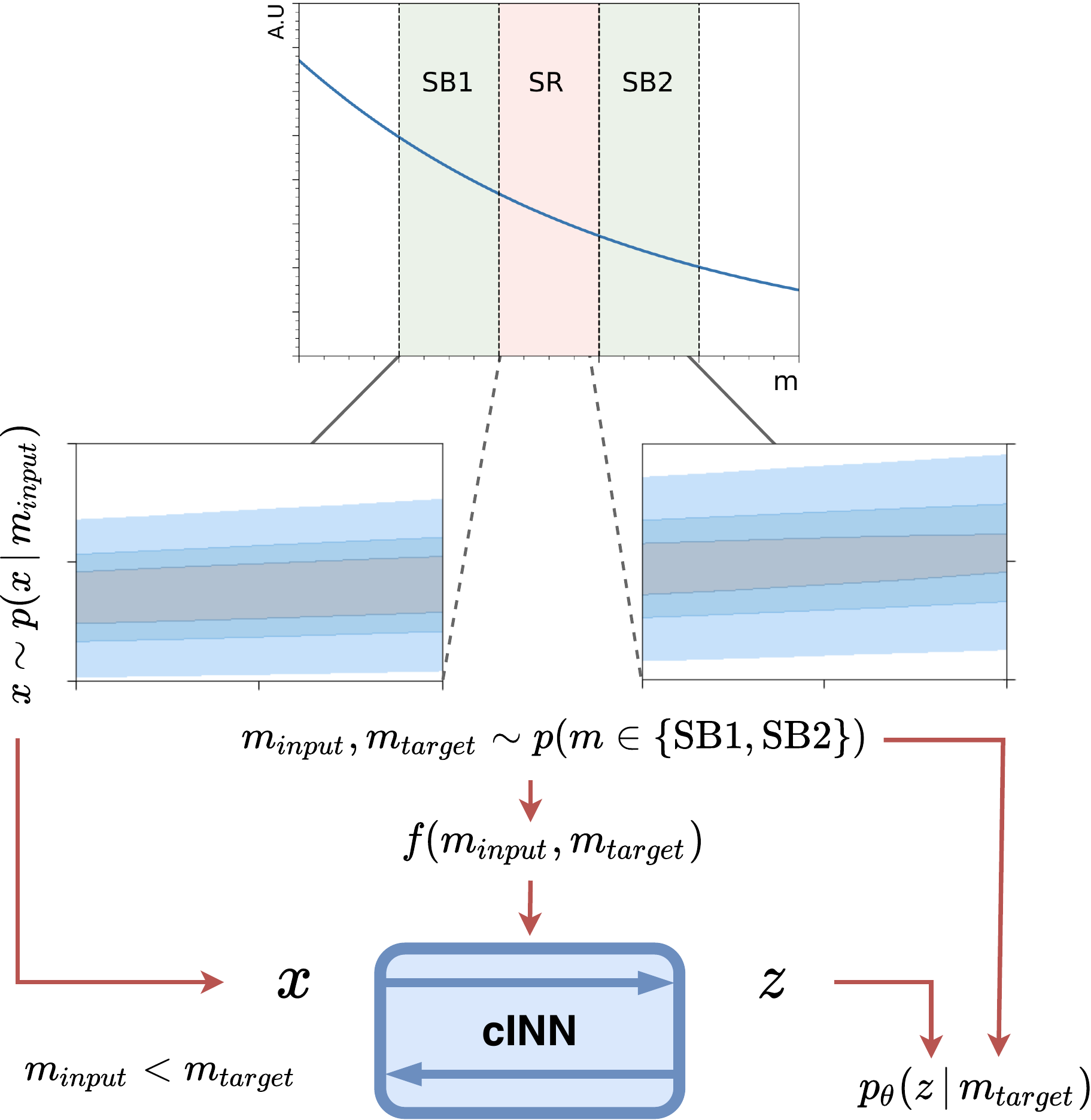}
    \caption{A schematic overview of the training procedure for \FfF for an event where the target \mjj value is greater than the input value.
        A single conditional normalizing flow is used for the base distribution, conditioned on the target \mjj value $m_\textit{target}$, to determine $p_\theta(z|m_\textit{target})$.
        The top normalizing flow is conditioned on a function of the input ($m_\textit{input}$) and target ($m_\textit{target}$) \mjj values.
        For the case where $m_\textit{target} < m_\textit{input}$, an inverse pass of the network is used and the conditioning property is calculated as $f(m_\textit{target} , m_\textit{input})$.}
    \label{fig:f4ftrain}
\end{figure}

The probability term is evaluated using a single base distribution trained on the data from \SBone and \SBtwo.
The loss for the batch is calculated from the average of the probabilities calculated from the forward and inverse passes.
A schematic overview is shown in \cref{fig:f4ftrain}.

\subsubsection*{Implementation}
The \FfF architecture comprises two conditional normalizing flows, the base distribution and the transformer flow.
The base distribution learns the conditional density of the training data which is used to train the top flow.
The top flow in turn transforms data from initial values of \mjj to target values.

The base distribution is trained on the side-band data with a standard normal distribution as the target prior.
It is conditioned on \mjj.
% It consists of eight autoregressive transformations using rational quadratic~(RQ) splines, defined by four bins.
The top flow is trained between data drawn from the side-bands.
The transformation is conditioned on $\Delta \mjj = m_{JJ}^{high} - m_{JJ}^{low}$.
The base distribution flow consists of ten autoregressive transformations using RQ splines, defined by four bins.
The top flow consists of eight coupling transformations using RQ splines, defined by four bins.
% Both the base distribution and top flow consists of eight autoregressive transformations using RQ splines, each defined by four bins.
They are trained separately using the Adam optimiser and a cosine annealing learning rate scheduler.
Each are trained for 100 epochs with a batch size of 256.

\subsection{Generating background samples}

To transform the data from the side-bands into the signal region, we assign target \mjj values corresponding to the SR to the data in each side-band.
The target values $m_\textit{target}$ are drawn from a function of the form
\begin{equation}
    f(z) = p_1 \left(1 - z \right)^{p_2} z^{p_3},
\end{equation}
where $z=\mjj/\sqrt{s}$, with parameters $p_i$ learned by performing a fit to the side-band data.
Data from \SBone (\SBtwo) are transformed in a forwards (inverse) pass through the top flow with $\Delta\mjj$.

The background template can be oversampled by assigning multiple target \mjj values to each data point.
This has been found to improve the performance of \CWoLa classifiers.

Due to the bidirectional nature of the cINNs, it is also possible to generate validation samples in regions further away from the SR.
These outer-bands can be used to optimise the hyperparameters of the top flow in \FfF.

\subsection{Comparison to \CURTAINs}

\FfF has a much simpler training procedure than in the original \CURTAINs.

% , as it removes the need for splitting the SBs and alternating between training \CURTAINs between \SBone and \SBtwo, and within each side-band.

In \CURTAINs, the Sinkhorn loss~\cite{cuturi2013sinkhorn} was used to train the network, with the distance measured between a batch of data sampled from the target region and the transformed data.
The target \mjj values for the transformed data were chosen to match the values in the target batch.
However, there was no guarantee that the minimum distance corresponded to the pairing of the transformed event with the event in the target batch with the corresponding $m_\textit{target}$ value.
Furthermore, the loss itself ignored the \mjj values as the input data and target data in the batch with the corresponding \mjj target value are not necessarily events that should look the same for the same \mjj value.
Although successful, this approach introduced a lot of stochasticity, and required a large number of epochs to converge.

Due to the new loss, the training in \FfF no longer needs to be between two discrete regions.
This has the added benefit that it removes the need for splitting the SBs and alternating between training \CURTAINs between \SBone and \SBtwo, and within each side-band.

Finally, in \CURTAINs the network was trained and updated alternating batches in the forward and inverse directions.
In \FfF a single batch has both increasing and decreasing target \mjj values. As such the network weights are updated based on the  average of the loss in both the forward and inverse directions for each individual batch.

Due to the additional base distribution, \FfF is no longer defined by a single model trained for each SR.
This introduces an extra model which needs to be trained and optimised.
We observe, however, that overall training both the base distribution and normalizing flow between SB data is less than required to train the cINN in \CURTAINs.

The additional time required to train the base distribution can also be minimised.
When training \FfF for multiple SRs, a single base distribution can be trained using all available data for all possible \mjj values.
For each SR, the network would only be evaluated for values in \SBone and \SBtwo, and no bias would be introduced from data in the SR.
This reduces the overall computational cost incurred when evaluating multiple signal regions.
\subsection{Comparison to other approaches}
This approach is one of several using normalizing flows as density models for background estimation for extending the sensitivity of bump hunts.
Many of these methods, including \CURTAINs and \FfF, produce background samples for use with \CWoLa bump hunting~\cite{cwola,cwolabump}.
In \CWoLa bump hunting, classifiers are trained on data from a hypothesised signal enriched region (the SR) and a signal depleted region (the SBs).
Cuts are applied onto the classifier score to reduce the amount of background and, in the presence of signal, enhance the sensitivity.

\begin{itemize}
    \item In \textsc{Anode}~\cite{anode}, conditional normalizing flows are trained to learn the probability of the signal and background from data drawn from the SBs and SR respectively.
    The normalizing flows are conditioned on \mjj, and the ratio of the probabilities is used as a likelihood test.

    \item In \CATHODE~\cite{cathode}, a conditional normalizing flow is trained on all SB data conditioned on \mjj.
    Samples with \mjj corresponding to the distribution of data in the SR (extrapolated from a side-band fit in \mjj) are generated using the normalizing flow.
    These generated samples form a synthetic background sample which together with the SR data are used in a \CWoLa approach~\cite{cwola,cwolabump}.

    \item In \FETA~\cite{feta}, the \emph{Flows for Flows} approach is used to train a conditional normalizing flow between background data in a monte carlo (MC) simulated sample and the data in the side-bands, as a function of \mjj.
    This flow is used to transport the MC events from the SR to the data space, and account for mismodelling observed in the simulated distributions.
    A \CWoLa classifier is trained on the transported MC and the SR data.
    
    \item \textsc{LaCathode}~\cite{Hallin:2022eoq} addresses the problem of distribution sculpting resulting from the choice of input features.
    Here the \CWoLa classifier is trained on the base density of \CATHODE, by first passing the SR data through the \CATHODE conditional normalizing flow and using samples drawn from the prior base distribution.
    This approach is complementary to any method training a conditional normalizing flow, such as \FfF and \FETA.

    \item Although not applied in the context of bump hunt searches, \textsc{ABCDnn}~\cite{Choi:2020bnf} uses normalizing flows to extrapolate data from one region to another, similar to \FETA. However, here the flows are trained with the maximum mean discrepancy loss, similar to the approach in \CURTAINs though it does not interpolate to unknown conditional values.

\end{itemize}

\section{Results}
The main measure of performance for background estimation approaches is by how much they improve the sensitivity to a signal in a \CWoLa bump hunt~\cite{cwolabump}.

We define a SR centred on the signal process with a width of 400~GeV, which contains nearly all of the signal events.
For \FfF and \CURTAINs, we use side-bands 200~GeV either side of the SR to train the methods.
Only these regions are used to train the base distribution for \FfF.
For \CATHODE, the whole \mjj distribution either side of the SR is considered as the SB region.
This corresponds to side-bands of widths 500~GeV and 900~GeV.
% This is found to improve the performance of the classifiers.

Weakly supervised classifiers are trained to separate the generated background samples from data in the SR.
% For \CURTAINs, \FfF and \CATHODE the background samples are oversampled with respect to the data in the SR by a factor of four.
For \CURTAINs, \FfF, and \CATHODE, an oversampling factor of four is used to generate the backround samples in the SR, at which point the performance reaches saturation.
In \CURTAINs and \FfF this is with respect to the transported SB data, whereas for \CATHODE it is based on the yields in the SR.
% \subsection{Generating background samples}

% To transform the data from the side-bands into the signal region, we assign target \mjj values corresponding to the SR to the data in each side-band.
% The \mjj values are drawn from a function of the form
% \begin{equation}
%     f(z) = p_1 \left(1 - z \right)^{p_2} z^{p_3},
% \end{equation}
% whose parameters are learned by performing a fit to the side-bands.
% Data from \SBone (\SBtwo) are transformed in a forwards (inverse) pass through the top flow.

% The background template can be oversampled by assigning multiple target \mjj values to each data point.
% This has been found to improve the performance of \CWoLa classifiers.
% For \CATHODE, data are sampled from the normalizing flow for the same \mjj values.
% For all three methods, a factor of four as many background samples are created as there are events in the SR in a process called \emph{oversampling}.
% This is found to improve the performance of the classifiers.

As a reference, a fully supervised classifier trained to separate the signal from background in this region, and an idealised classifier trained with a perfect background estimation are also shown.
The idealised classifier is trained for both equal numbers of background in each class (Eq-Idealised) and an oversampled background (Over-Idealised).

A $k$-fold training strategy with five folds is employed to train all classifiers.
Three fifths are used to train the classifier, with one fifth for validation and the final fifth as a hold out set.
The classifiers comprise three hidden layers with 32 nodes and ReLU activations.
They are trained for 20 epochs with the Adam optimiser and a batch size of 128.
The initial learning rate is $10^{-4}$ but is annealed to zero following a cosine schedule.

% Different amounts of signal are injected into the training sample in order to test the performance over a wide range of signal fractions.

\subsection{Comparison of performance}

\begin{figure}[hbpt]
    \centering
    \begin{subfigure}{0.48\textwidth}
        \centering
        \includegraphics[width=\textwidth]{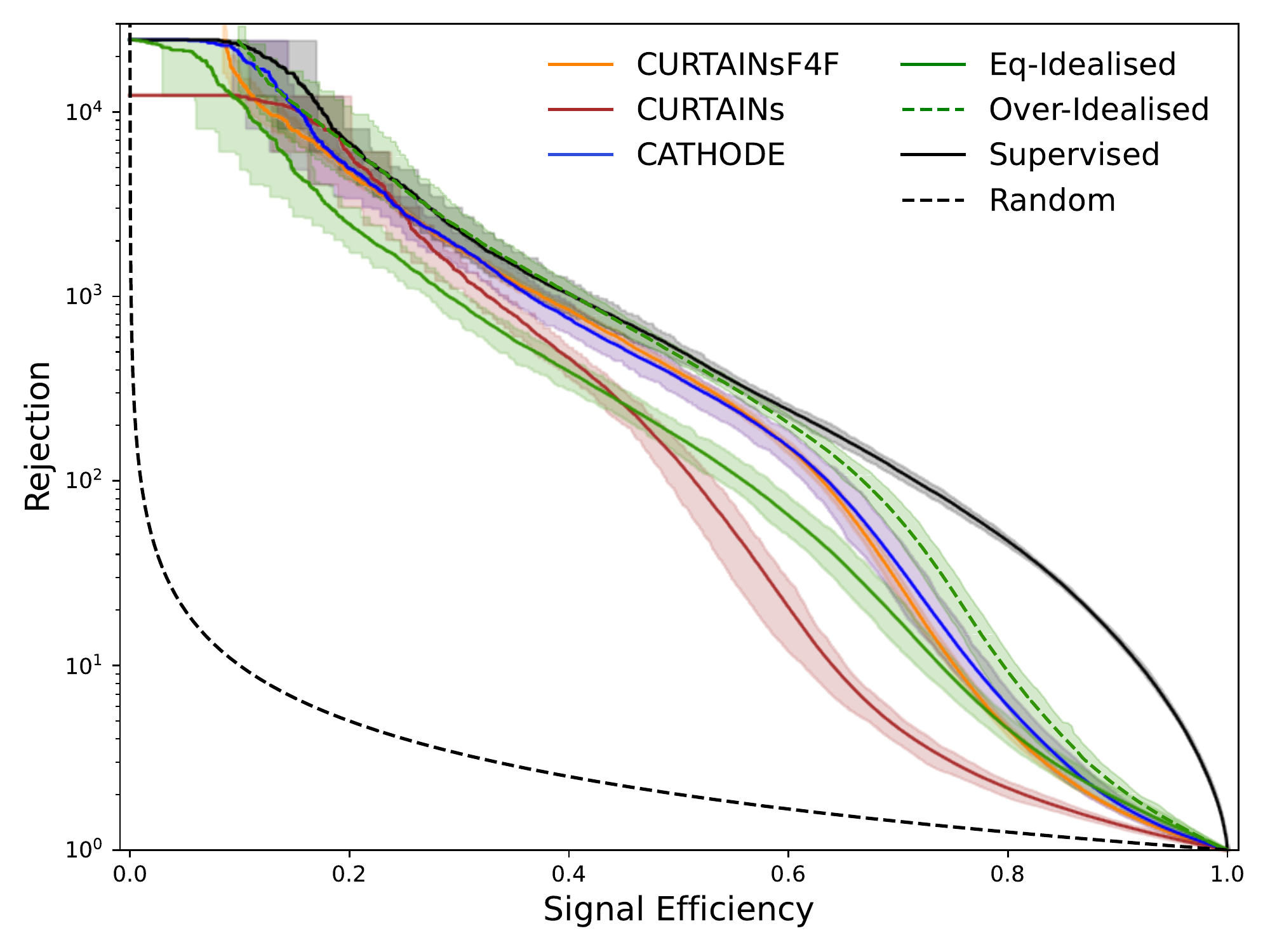}
    \end{subfigure}
    \hfill
    \begin{subfigure}{0.48\textwidth}
        \centering
        \includegraphics[width=\textwidth]{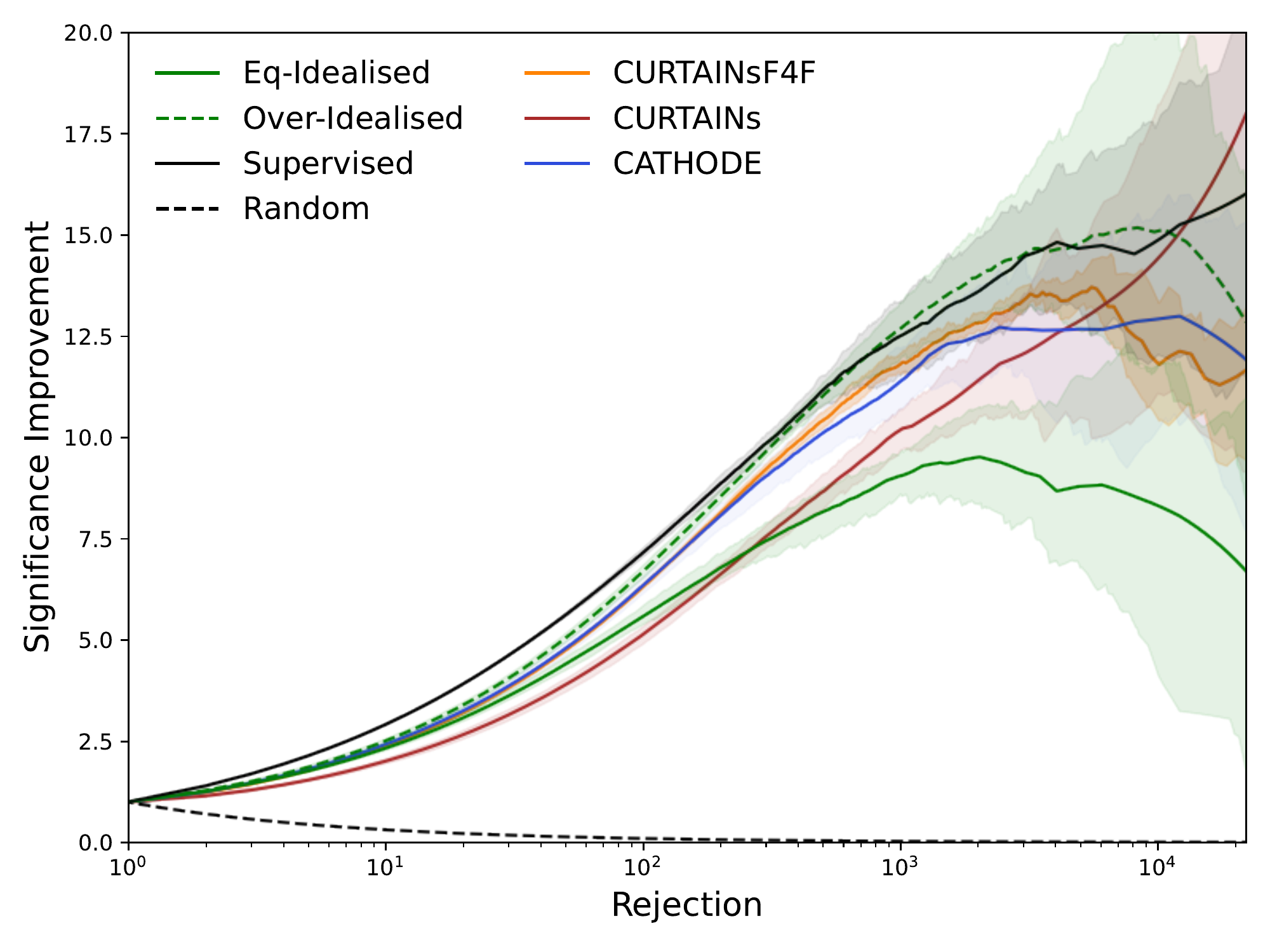}
    \end{subfigure}
    \caption{Background rejection as a function of signal efficiency (left) and significance improvement as a function of background rejection (right) for \CURTAINs~(red), \FfF~(orange), \CATHODE~(blue), Supervised~(black), Eq-Idealised~(green, solid), and Over-Idealised~(green, dashed).
        All classifiers are trained on the sample with 3,000 injected signal events,
        using a signal region $3300\leq\mjj<3700$~GeV.
        The lines show the mean value of fifty classifier trainings with different random seeds with the shaded band covering 68\% uncertainty.
        A supervised classifier and two idealised classifiers are shown for reference.
    }
    \label{fig:sic_roc}
\end{figure}

\Cref{fig:sic_roc} shows the background rejection and significance improvement for \CWoLa classifiers trained using the different background estimation models as the cut on the classifier is varied.
Here 3,000 signal events have been added to the QCD dijet sample, of which 2,214 fall within the SR.
\FfF shows significant improvement over the original \CURTAINs method, and now matches the \CATHODE performance across the majority of rejection and signal efficiency values.
This is despite being trained on a much smaller range of data.
\CURTAINs still displays better significant improvement at very high rejection values, however this is in a region dominated by the statistical uncertainty.

\begin{figure}[hbpt]
    \centering
    \begin{subfigure}{0.49\textwidth}
        \centering
        \includegraphics[width=\textwidth]{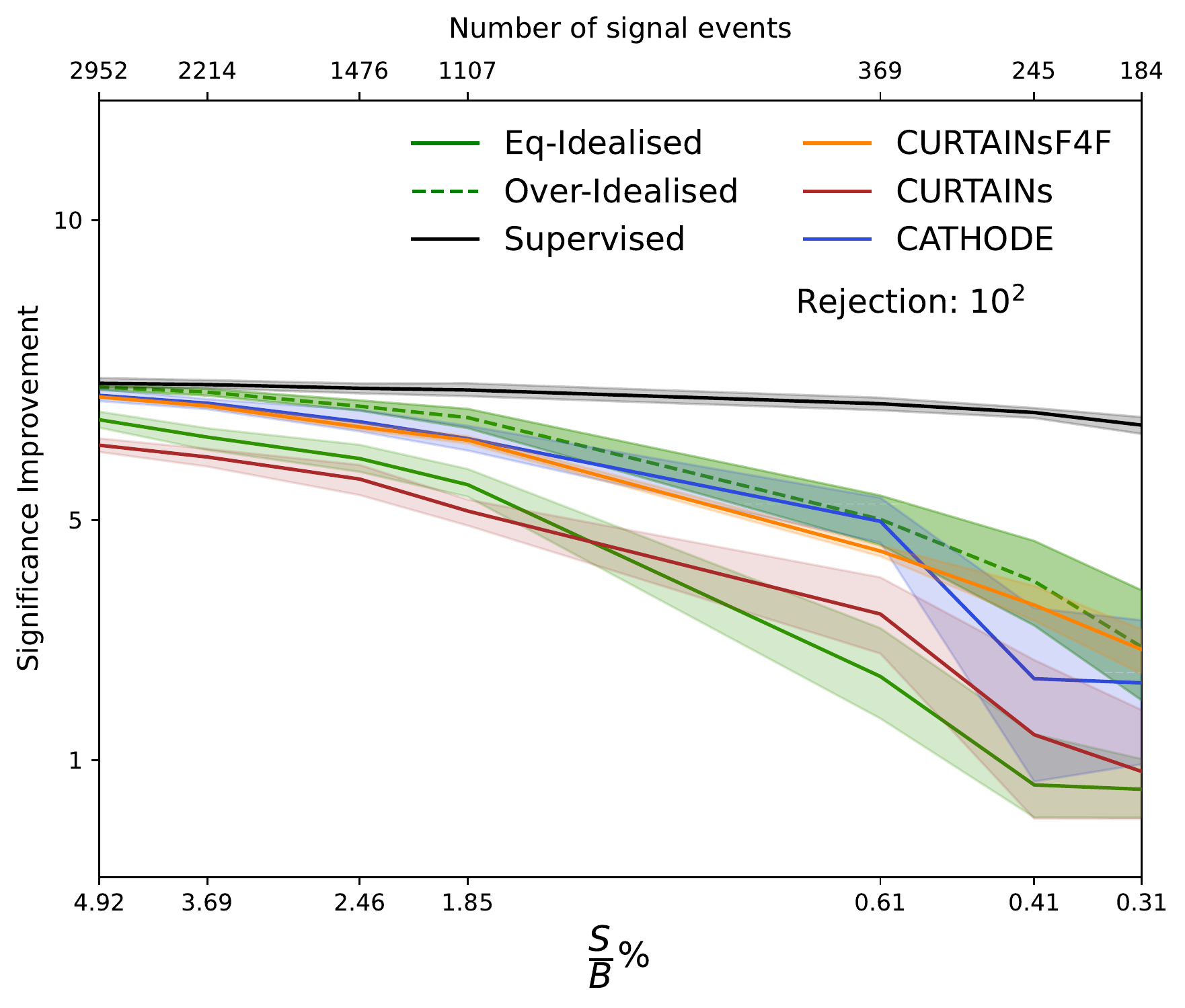}
    \end{subfigure}
    \begin{subfigure}{0.49\textwidth}
        \centering
        \includegraphics[width=\textwidth]{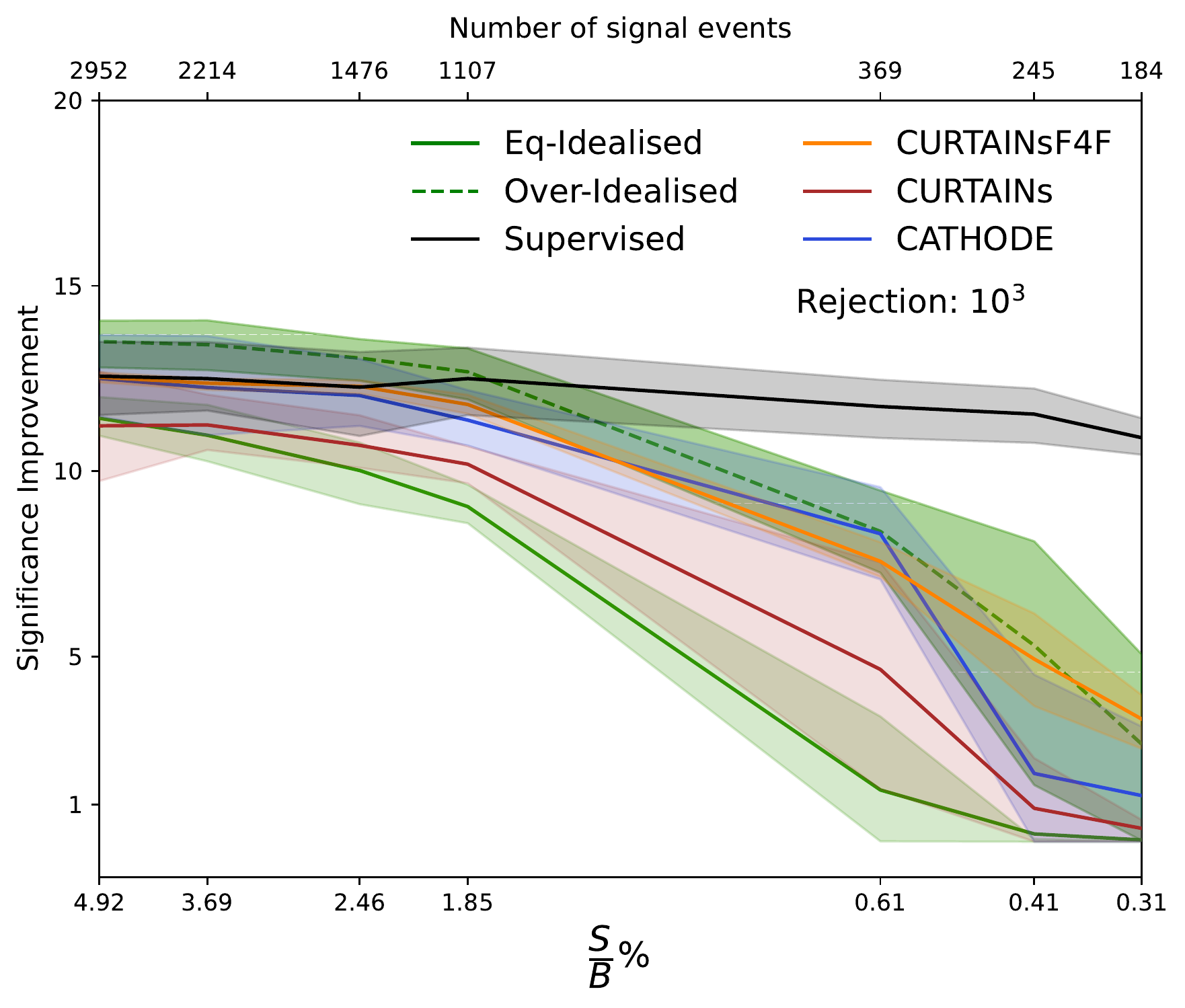}
    \end{subfigure}
    \caption{Significance improvement at a background rejection of $10^2$ (left) and $10^3$ (right) as a function of signal events in the signal region $3300\leq\mjj<3700$~GeV,  for \FfF~(orange), \CURTAINs~(red), \CATHODE~(blue), Supervised~(black), Eq-Idealised~(green, solid), and Over-Idealised~(green, dashed).
        The lines show the mean value of fifty classifier trainings with different random seeds with the shaded band covering 68\% uncertainty.
        A supervised classifier and two idealised classifiers are shown for reference.
    }
    \label{fig:sic_vs_sig}
\end{figure}

In \cref{fig:sic_vs_sig} the significance improvement for each method is calculated as a function of the number of signal events added to the sample.
Here both the signal fraction and raw number of signal events in the SR are reported.
The significance improvement is shown for two fixed background rejection values, rather than the maximum significance improvement, due to the sensitivity to fluctuations in the high background rejection regions where there are much lower statistics.
The performance of \FfF is improved across all levels of signal in comparison to the original \CURTAINs method, and performs equally as well as \CATHODE.

\subsection{Dependence on side-band width}

\begin{figure}[htbp]
    \centering
    \begin{subfigure}{0.48\textwidth}
        \centering
        \includegraphics[width=\textwidth]{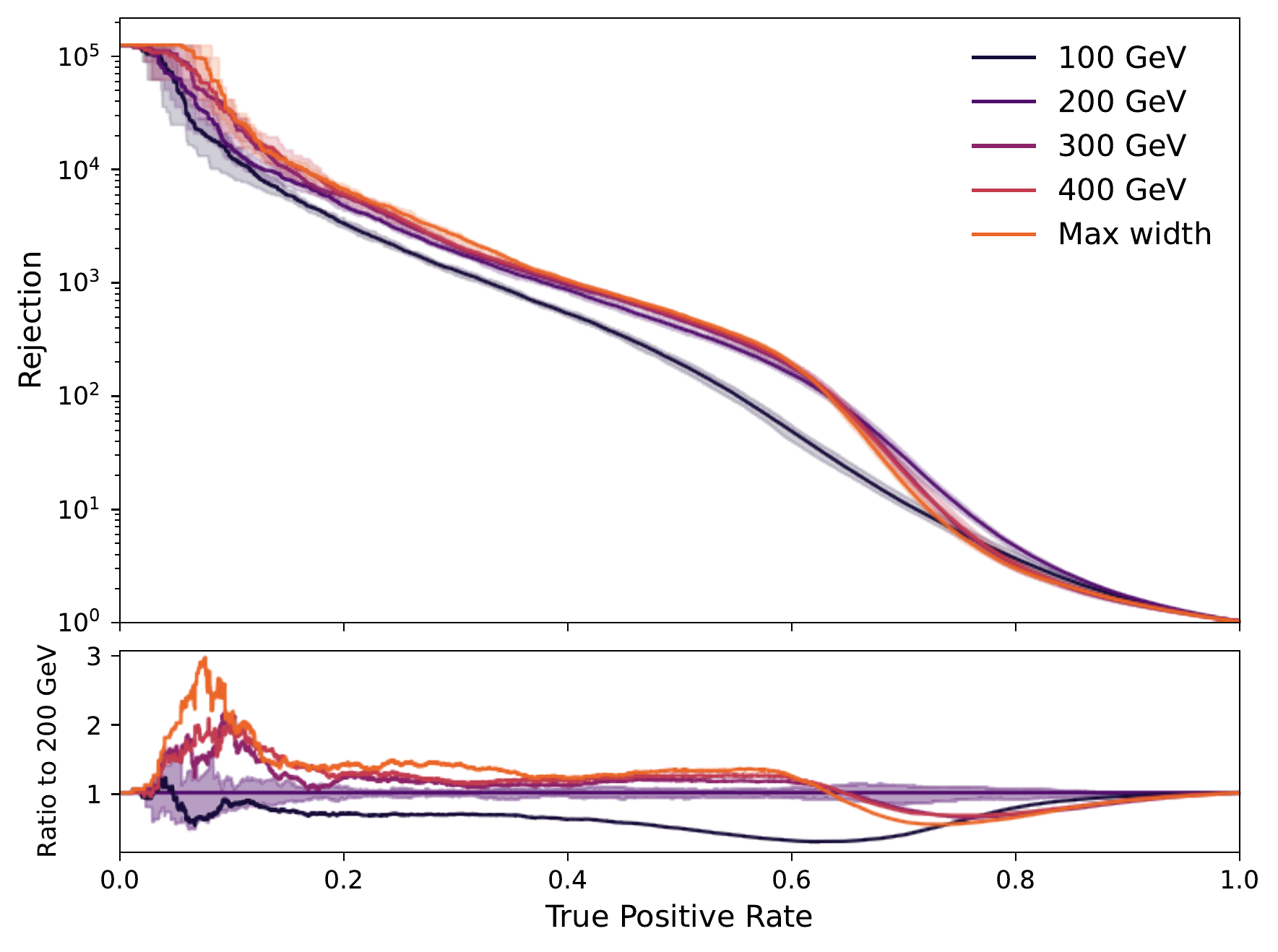}
    \end{subfigure}
    \hfill
    \begin{subfigure}{0.48\textwidth}
        \centering
        \includegraphics[width=\textwidth]{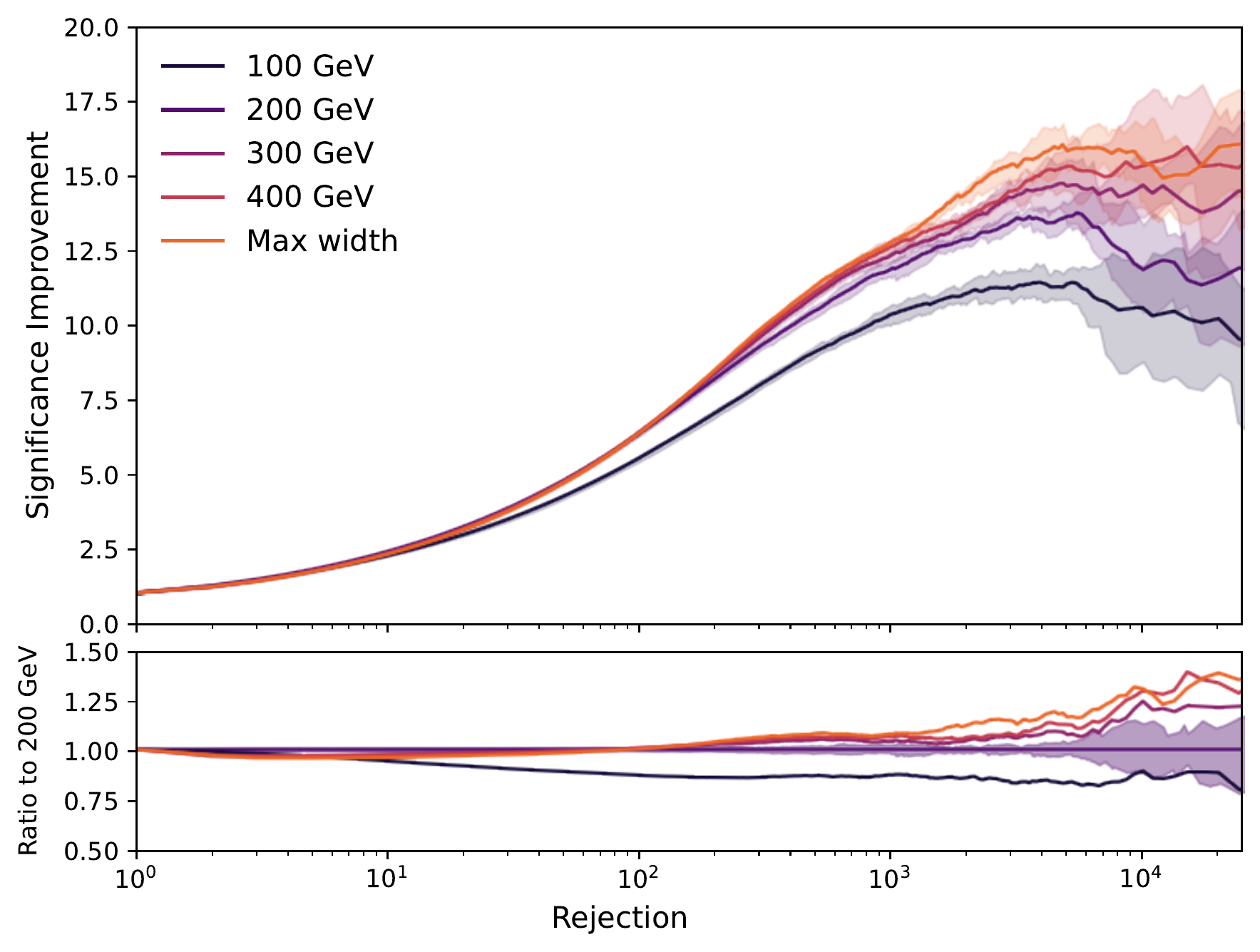}
    \end{subfigure}
    \caption{Background rejection as a function of signal efficiency (left) and significance improvement as a function of background rejection (right) for \FfF trained with varying width side-bands, ranging from 100~GeV to the maximum width possible (\SBone: 500~GeV, \SBtwo: 900~GeV).
        All classifiers are trained on the sample with 3,000 injected signal events,
        using a signal region $3300\leq\mjj<3700$~GeV.
        The lines show the mean value of fifty classifier trainings with different random seeds with the shaded band covering 68\% uncertainty.
    }
    \label{fig:f4fside-bands}
\end{figure}

In \FfF, 200 GeV wide side-bands are used to train the networks and learn a local transformation.
With leakage of signal into the side-bands or changing background composition, it could be beneficial to have narrower or wider SBs, and there is no set prescription for which is optimal.
\Cref{fig:f4fside-bands} shows the impact on performance of varying the widths from 100~GeV up to all data not contained in the signal region (max width).
For the 100 GeV wide side-bands a noticeable drop in performance is observed in the significance improvement and ROC curves.
However at a background rejection of $\sim 10^{3}$ all other side-band widths have similar levels of rejection.
At higher levels of background rejection training on larger side-bands, and thus more data, results in better performance than the default \FfF model with widths of 200~GeV.
It should be kept in mind that as the width of the side-bands increase, the required training time increases.
For these comparisons no hyperparameter optimisation has been performed and the default values are used for all models.

\begin{figure}[hbpt]
    \centering
    \begin{subfigure}{0.48\textwidth}
        \centering
        \includegraphics[width=\textwidth]{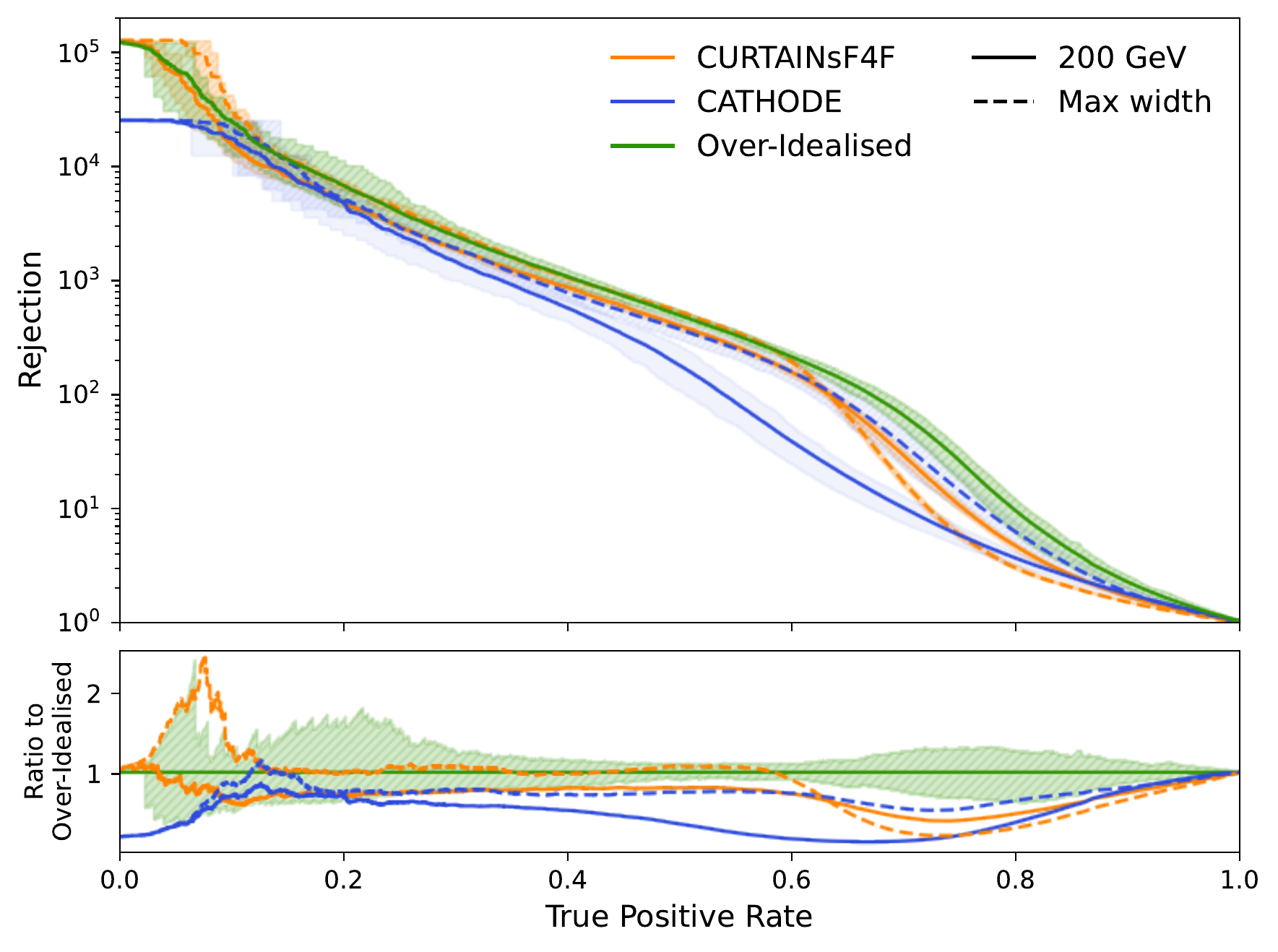}
    \end{subfigure}
    \hfill
    \begin{subfigure}{0.48\textwidth}
        \centering
        \includegraphics[width=\textwidth]{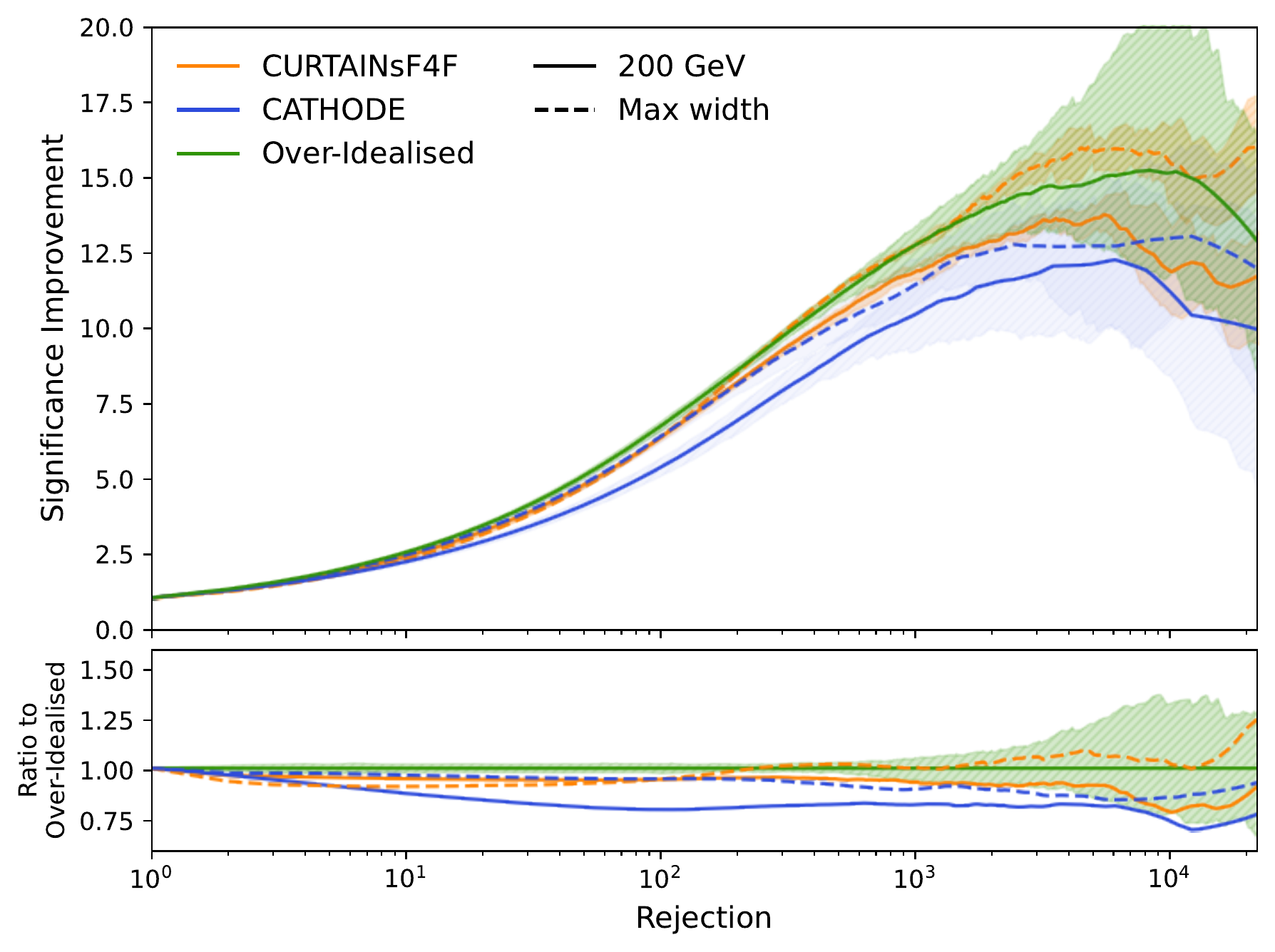}
    \end{subfigure}
    \caption{Background rejection as a function of signal efficiency (left) and significance improvement as a function of background rejection (right) for \FfF~(orange) and \CATHODE~(blue).
        Two side-band widths are used to train the two methods, 200~GeV side-bands (solid) and the maximum width (dashed, \SBone: 500~GeV, \SBtwo: 900~GeV).
        All classifiers are trained on the sample with 3,000 injected signal events,
        using a signal region $3300\leq\mjj<3700$~GeV.
        The lines show the mean value of fifty classifier trainings with different random seeds with the shaded band covering 68\% uncertainty.
        The Over-Idealised classifier (green) is shown for reference.
    }
    \label{fig:200_v_max_FfF_cath}
\end{figure}

In \cref{fig:200_v_max_FfF_cath} the performance of \FfF and \CATHODE are shown for the case where each model is trained on either 200~GeV wide SBs or the max width.
For \FfF the difference in performance is mostly at high background rejection whereas \CATHODE has a drop in performance at all values.

\subsection{Required training time}

For a bump hunt or sliding window search, a large number of models need to be trained which can result in a high demand on computing resources.
As a result, the granularity of a search may be restricted in line with overall computational time.
Therefore, a key measure of methods like \FfF and \CATHODE is how quick the models are to train.

In \cref{tab:trainingtime} the require time to train the two approaches for one SR are shown for convergence and for one epoch.
\CATHODE has an advantage over \FfF in that only one normalizing flow is trained.
The total training time required for \FfF is much reduced in comparison to \CURTAINs and is slightly faster than \CATHODE for the default configurations.

\begin{table}[htbp]
    \caption{Comparison of the required time to train \CURTAINs, \FfF, and \CATHODE.
        All models are trained on the same hardware with epoch and total training time representative of using an \nvidia~RTX~3080 graphics card.
        For \FfF two numbers are shown for the epoch time and number of epochs due to the two normalizing flows which need to be trained.
        Default side-band widths are used for all models, around the nominal signal region.
    }
    \label{tab:trainingtime}
    \centering
    \begin{tabularx}{0.8\textwidth}{R  Y Y  Y}
        \toprule
                  & Time / epoch [s] & $N$ epochs & Total time [min] \\
        \midrule
        \CURTAINs & 10               & 1000       & 167              \\
        \CATHODE  & 78               & 100        & 129              \\
        \FfF      & 32/32            & 100/100    & 107              \\
        \bottomrule
    \end{tabularx}
\end{table}

\subsection{Reducing computational footprint}

When applying the models to multiple signal regions in a bump hunt, new models need to be trained for each step.
For \CATHODE this involves training a complete model each time.
However, due to the modular nature of \FfF, if the base distribution is trained on the whole spectrum, only the top flow needs to be trained with each step.
As such, as soon as more than one SR is considered, \FfF requires substantially less computational resources for a similar level of performance.

Additionally, the transformation learned by the top flow in \FfF is known to be a smaller shift than for the base distribution or in \CATHODE.
The top flow can thus also be optimised for speed without sacrificing as much performance and does not require the same expressive architecture as used by default.
% By initialising the normalizing flow to the identity, fewer training epochs are expected to be required to reach convergence.

% The default \FfF configuration is compared to an efficient implementation in which a single base distribution is trained on all data, and the top flow is optimised for speed.
The default \FfF configuration is compared to an efficient implementation in which a single base distribution is trained on all data, and the top flow is optimised for speed.
The base distribution has the same architecture as the default configuration.
The efficient top flow comprises two coupling transformations using RQ splines, rather than eight, with each now defined by six bins instead of four.
% Larger networks used to predict the bins positions for the splines than in the default configuration.
% The transformation is initialised to the identity and trained for 20 epochs with a batch size of 256.
The top flow is trained for 20 epochs with a batch size of 256.
All other hyperparameters remain unchanged and side-bands of 200~GeV are used to train the top flow and produce the background template in the SR.
The potential reduction in computation time for using \FfF in a sliding window search is presented in \cref{tab:fasttrainingtime}.
With the efficient configuration more than one hundred signal regions can be evaluated with \FfF transformers for the same computational cost as ten with the default configuration.

\begin{figure}[hbpt]
    \centering
    \begin{subfigure}{0.48\textwidth}
        \centering
        \includegraphics[width=\textwidth]{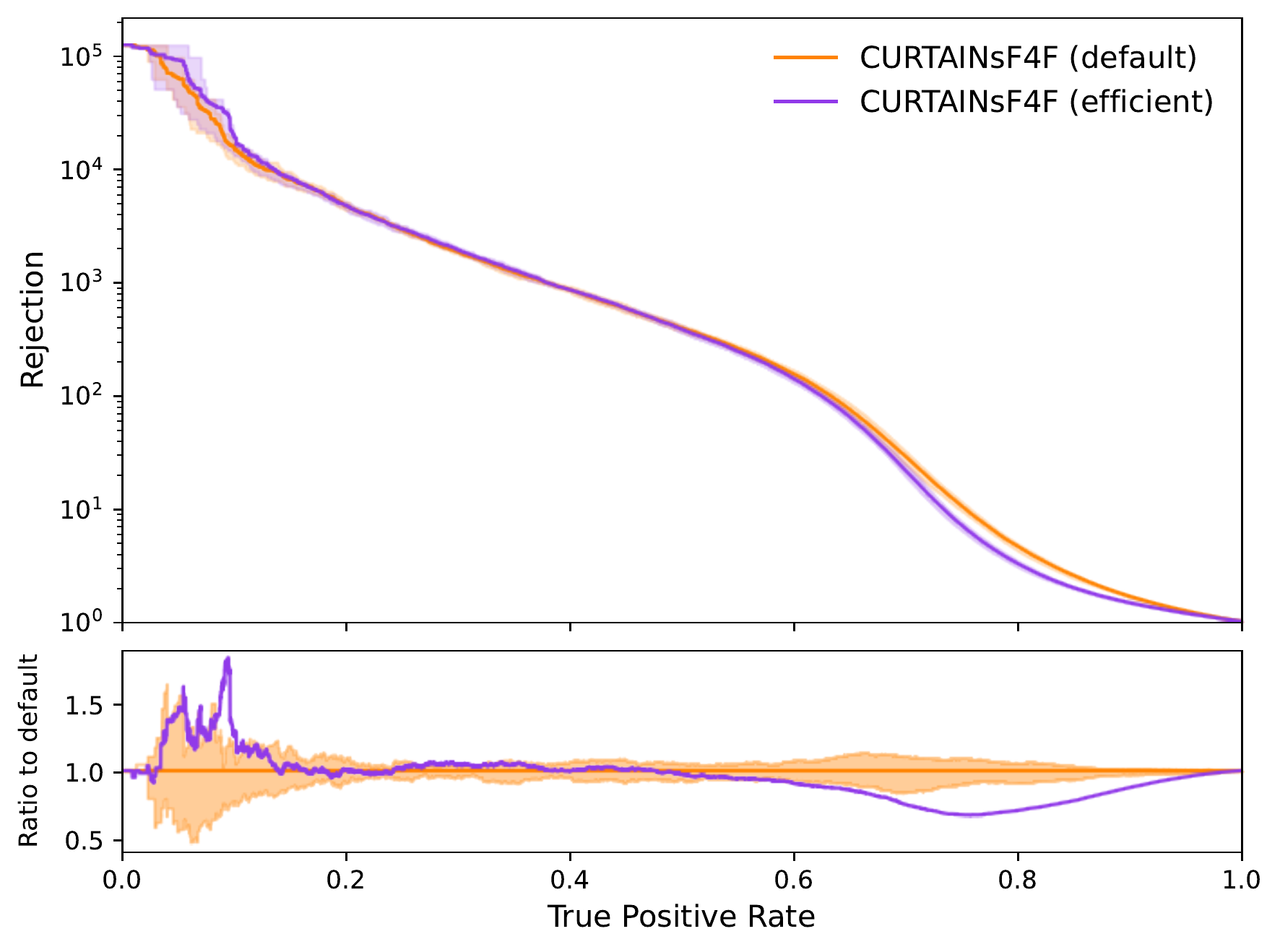}
    \end{subfigure}
    \hfill
    \begin{subfigure}{0.48\textwidth}
        \centering
        \includegraphics[width=\textwidth]{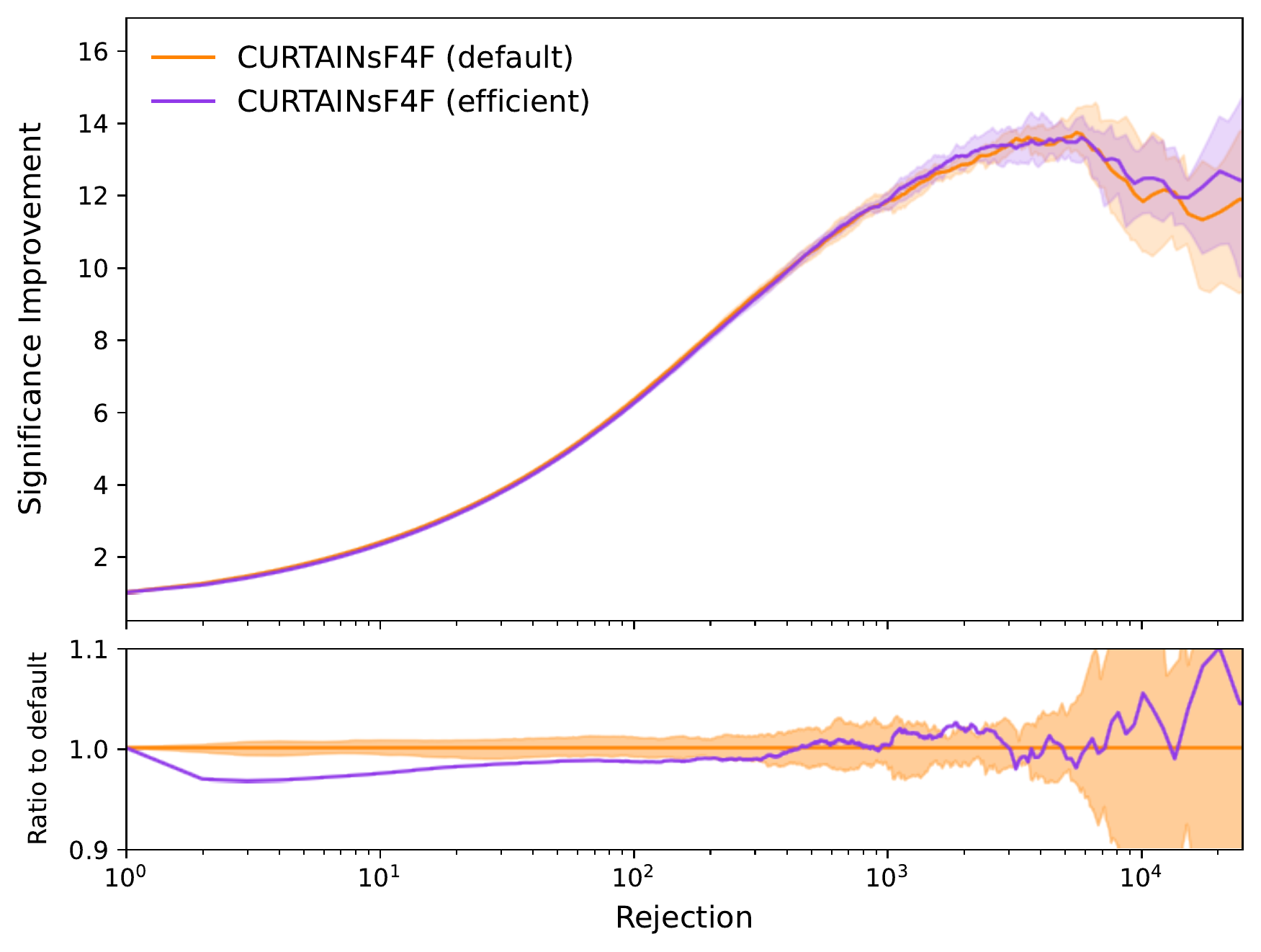}
    \end{subfigure}
    \caption{Background rejection as a function of signal efficiency (left) and significance improvement as a function of background rejection (right) for \FfF using the default~(orange) and efficient~(purple) training configurations.
        All classifiers are trained on the sample with 3,000 injected signal events,
        using a signal region $3300\leq\mjj<3700$~GeV.
        The lines show the mean value of fifty classifier trainings with different random seeds with the shaded band covering 68\% uncertainty.
        % A supervised classifier and two idealised classifiers are shown for reference.
    }
    \label{fig:fasttrainingcomp}
\end{figure}

\begin{figure}[hbpt]
    \centering
    % \begin{subfigure}{0.5\textwidth}
        % \centering
        \includegraphics[width=0.65\textwidth]{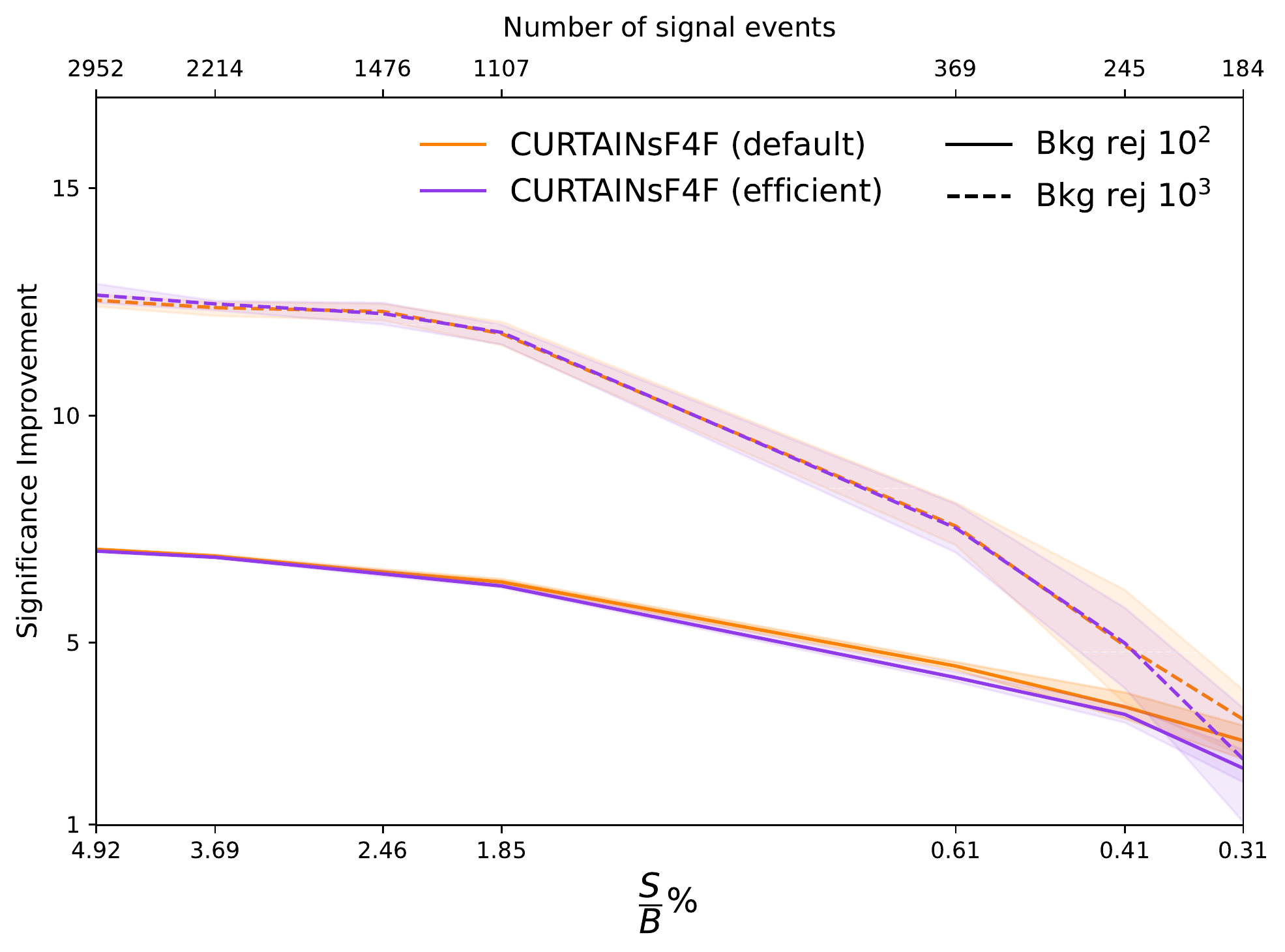}
        % \includegraphics[width=\textwidth]{figures/roc_3000}
    % \end{subfigure}
    % \begin{subfigure}{0.49\textwidth}
    %     \centering
    %     \includegraphics[width=\textwidth]{figures/sic_at1E3_defaultveff.pdf}
    %     % \includegraphics[width=\textwidth]{figures/roc_3000}
    % \end{subfigure}
    \caption{Significance improvement at a background rejection of $10^2$ and $10^3$ as a function of signal events in the signal region $3300\leq\mjj<3700$~GeV for \FfF using the default~(orange) and efficient~(purple) training configurations.
        The lines show the mean value of fifty classifier trainings with different random seeds with the shaded band covering 68\% uncertainty.
        % A supervised classifier and two idealised classifiers are shown for reference.
    }
    \label{fig:fasttrainingnsig}
\end{figure}

In \cref{fig:fasttrainingcomp} the significance improvement when using the efficient configuration is compared to the default \FfF model for 3000 injected signal events.
The performance as a function of the number of injected signal events is shown in \cref{fig:fasttrainingnsig}.
No significant decrease in performance is observed.

\begin{table}[htbp]
    \caption{Comparison of the required time to train the base distribution and top flow in \FfF.
        The default configuration comprises the base distribution and top flow trained on 200~GeV side-bands.
        The efficient configuration has a single base distribution trained on all data, and a top flow trained on 200~GeV side-bands and optimised for the fastest training time.
        All models are trained on the same hardware with epoch and total training time representative of using an \nvidia~RTX~3080 graphics card.
        An extrapolation of the required total time to train a complete \FfF model for one and ten signal regions are also shown for the two configurations.
        The extrapolated time for 125 signal regions is also shown for the efficient configuration, requiring less time than ten signal regions with the default configuration.\\
        $^\dagger$ Timing is for the nominal side-bands, this would vary as the signal region changes due to total number of training events.
        }
    \label{tab:fasttrainingtime}
    \centering
    \begin{tabularx}{0.8\textwidth}{R  Y Y  Y}
        \toprule
                                                                       & Time / epoch [s]        & $N$ epochs & Total time [min] \\
        \midrule
        \multicolumn{4}{c}{Default}                                                                                              \\
        \midrule
        Base                                                           & 32.4$^\dagger$          & 100        & 54               \\
        Top flow                                                       & 31.5$^\dagger$          & 100        & 53               \\
        \midrule
        \multicolumn{3}{r}{One Signal Region}                          & 107                                                     \\
        \multicolumn{3}{r}{(Extrapolated$^\dagger$) Ten Signal Region} & 1070                                                    \\
        \midrule
        % \multicolumn{4}{c}{Faster}                                                                                     \\
        % \midrule
        % Base                                                           & \textbf{32.4}$^\dagger$   & 100        & 54               \\
        % % \midrule
        % Top flow                                                       & \textbf{21.3}$^\dagger$   & 20         & 7                \\
        % \midrule
        % \multicolumn{3}{r}{One Signal Region}                          & 61                                              \\
        % \multicolumn{3}{r}{(Extrapolated$^\dagger$) Ten Signal Region} & 610                                              \\
        % \multicolumn{3}{r}{(Extrapolated$^\dagger$) 17 Signal Region} & 1037                                             \\
        % \midrule
        \multicolumn{4}{c}{Efficient}                                                                                            \\
        \midrule
        Base                                                           & 104.2                   & 100        & 174              \\
        % \midrule
        Top flow                                                       & 21.3$^\dagger$          & 20         & 7                \\
        \midrule
        \multicolumn{3}{r}{One Signal Region}                          & 181                                                     \\
        \multicolumn{3}{r}{(Extrapolated$^\dagger$) Ten Signal Region} & 244                                                     \\
        \multicolumn{3}{r}{(Extrapolated$^\dagger$) 125 Signal Region} & 1049                                                    \\
        \bottomrule
    \end{tabularx}
\end{table}

\FloatBarrier
\section{Conclusions}
In the original \CURTAINs method, a distance based optimal transport loss was used to train a conditional invertible neural network.
In this work we have shown that the performance can be improved significantly by moving to a maximum likelihood estimation loss, using the \emph{Flows for Flows} methodology.
The performance levels reached by \FfF are state-of-the-art, and can do so training on less data from narrower side-bands than the previous state of the art.
% Furthermore, the required time for training this method in comparison to the previous approach is reduced by more than half.
% Initialisation of the top flow to the identity could also bring even further reductions in the time required to train the networks.

By only modifying the training procedure, other advantages of \CURTAINs are preserved.
Additional validation regions further away from the signal region can be used to optimise the hyperparameters of both the normalizing flow and classification networks.

Furthermore, in order to address background sculpting resulting from the classifiers, the latent approach introduced in \textsc{La}\CATHODE can be performed using the base distribution.
With the original \CURTAINs method, an additional normalizing flow would need to be trained on the signal region data for each signal region.

Finally, for a single signal region \FfF requires similar computing resources as other leading approaches, with almost half the required training time in comparison to the original \CURTAINs method.
However, when moving to a sliding window bump hunt, the overall computing resources required for \FfF is reduced by a large factor.
On the LHCO R\&D dataset over one hundred signal regions can be trained for the same computing resources as otherwise required for ten signal regions.
This could be of particular interest for large scale searches which are limited by the computational cost to cover a larger number of signal regions, such as those in Refs.~\cite{Shih:2021kbt,Shih:2023jfv} amongst others.

% TODO: include author contributions
% \paragraph{Author contributions}
% This is optional. If desired, contributions should be succinctly described in a single short paragraph, using author initials.

% TODO: include funding information
\section*{Acknowledgements}
We would like to thank David Shih and Matt Buckley for valuable discussions at the ML4Jets 2022 conference at Rutgers, New Jersey, in particular on results of interest and potential applications.
In addition, we would like to thank Radha Mastandrea, Ben Nachman and Kees Benkendorfer for useful discussions concerning the performance evaluation.

The authors would like to acknowledge funding through the SNSF Sinergia grant called "Robust Deep Density Models for High-Energy Particle Physics and Solar Flare Analysis (RODEM)" with funding number CRSII$5\_193716$ and the SNSF project grant 200020\_212127 called "At the two upgrade frontiers: machine learning and the ITk Pixel detector".

% TODO:
% Provide your bibliography here. You have two options:

% FIRST OPTION - write your entries here directly, following the example below, including Author(s), Title, Journal Ref. with year in parentheses at the end, followed by the DOI number.
%\begin{thebibliography}{99}
%\bibitem{1931_Bethe_ZP_71} H. A. Bethe, {\it Zur Theorie der Metalle. i. Eigenwerte und Eigenfunktionen der linearen Atomkette}, Zeit. f{\"u}r Phys. {\bf 71}, 205 (1931), \doi{10.1007\%2FBF01341708}.
%\bibitem{arXiv:1108.2700} P. Ginsparg, {\it It was twenty years ago today... }, \url{http://arxiv.org/abs/1108.2700}.
%\end{thebibliography}

% \bibliographystyle{SciPost_bibstyle} % Include this style file here only if you are not using our template
\bibliography{bib/rodem.bib}

\clearpage
\appendix

\section{Additional results}

\begin{figure}[hbpt]
    \centering
    \begin{subfigure}{0.65\textwidth}
        \centering
        \includegraphics[width=\textwidth]{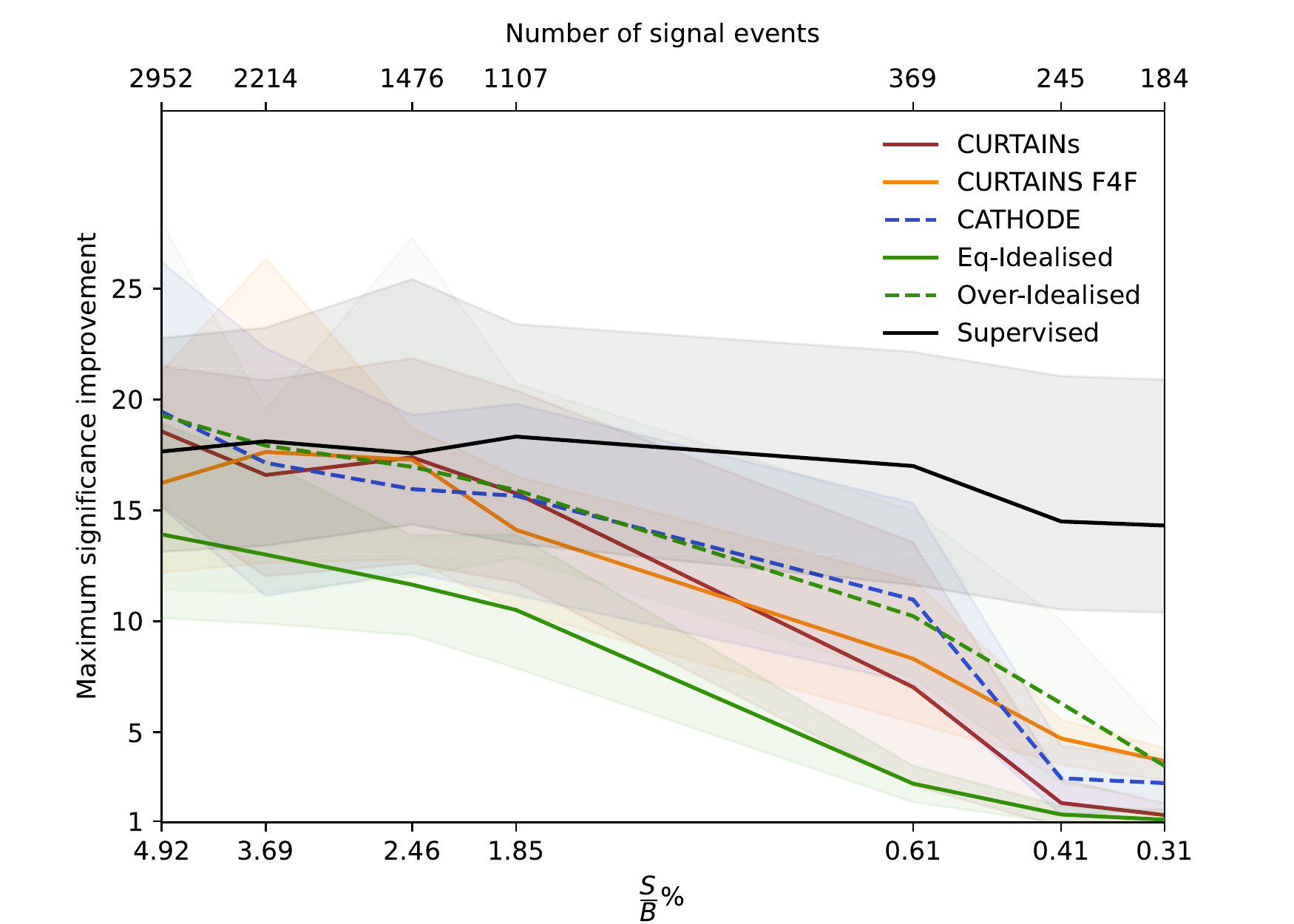}
    \end{subfigure}
    \caption{Maximum significance improvement as a function of signal events in the signal region $3300\leq\mjj<3700$~GeV,  for \CURTAINs~(red), \FfF~(orange), \CATHODE~(blue), Supervised~(black), Eq-Idealised~(green, solid), and Over-Idealised~(green, dashed).
        The lines show the mean value of fifty classifier trainings with different random seeds with the shaded band covering 68\% uncertainty.
        A supervised classifier and two idealised classifiers are shown for reference.
    }
    \label{fig:app:max_sic_vs_sig}
\end{figure}

In \cref{fig:app:max_sic_vs_sig} the maximum significance improvement for the default models is shown, rather than at fixed background rejection values.

\begin{figure}[htbp]
    \centering
    \begin{subfigure}{0.48\textwidth}
        \centering
        \includegraphics[width=\textwidth]{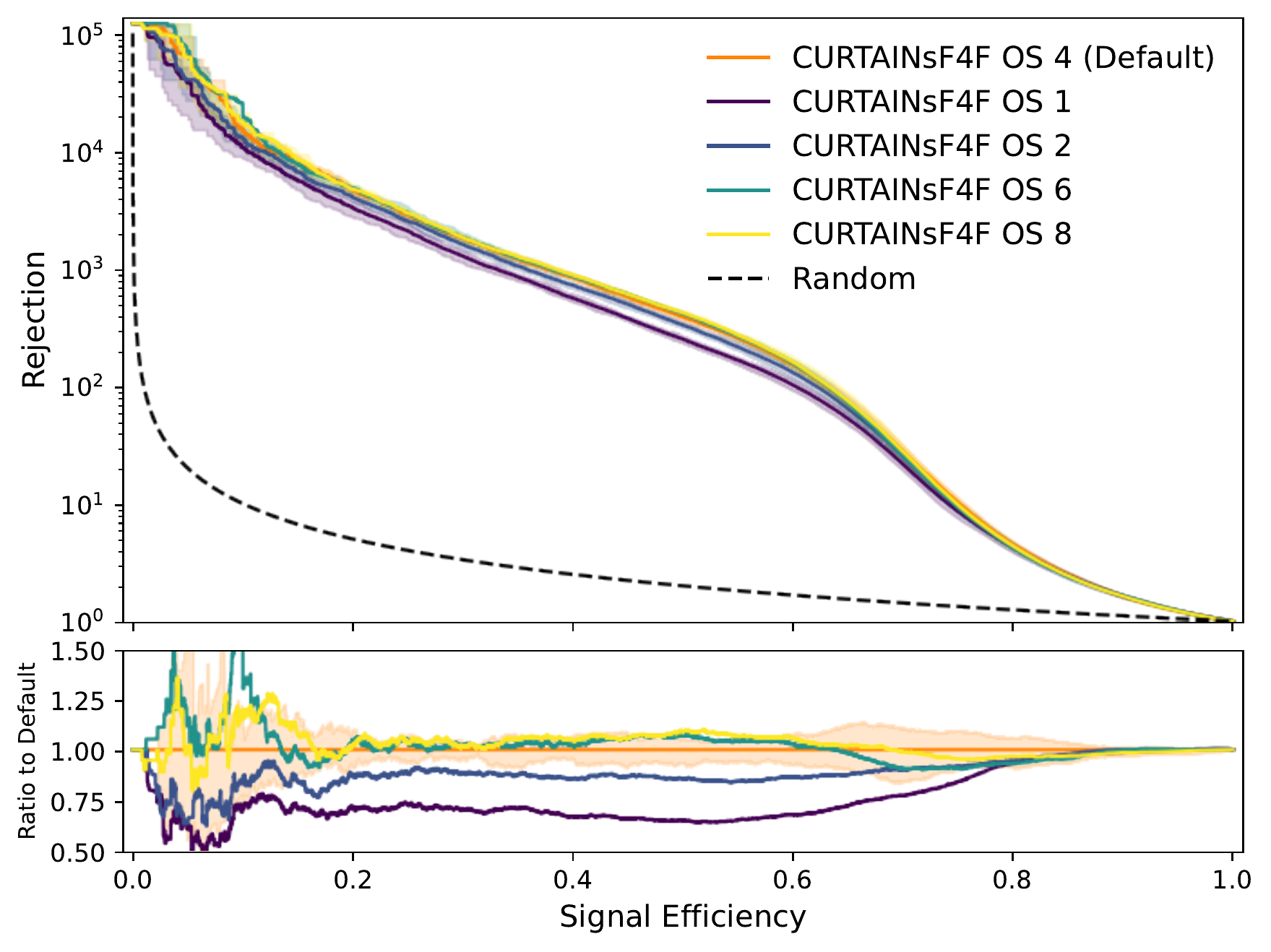}
    \end{subfigure}
    \hfill
    \begin{subfigure}{0.48\textwidth}
        \centering
        \includegraphics[width=\textwidth]{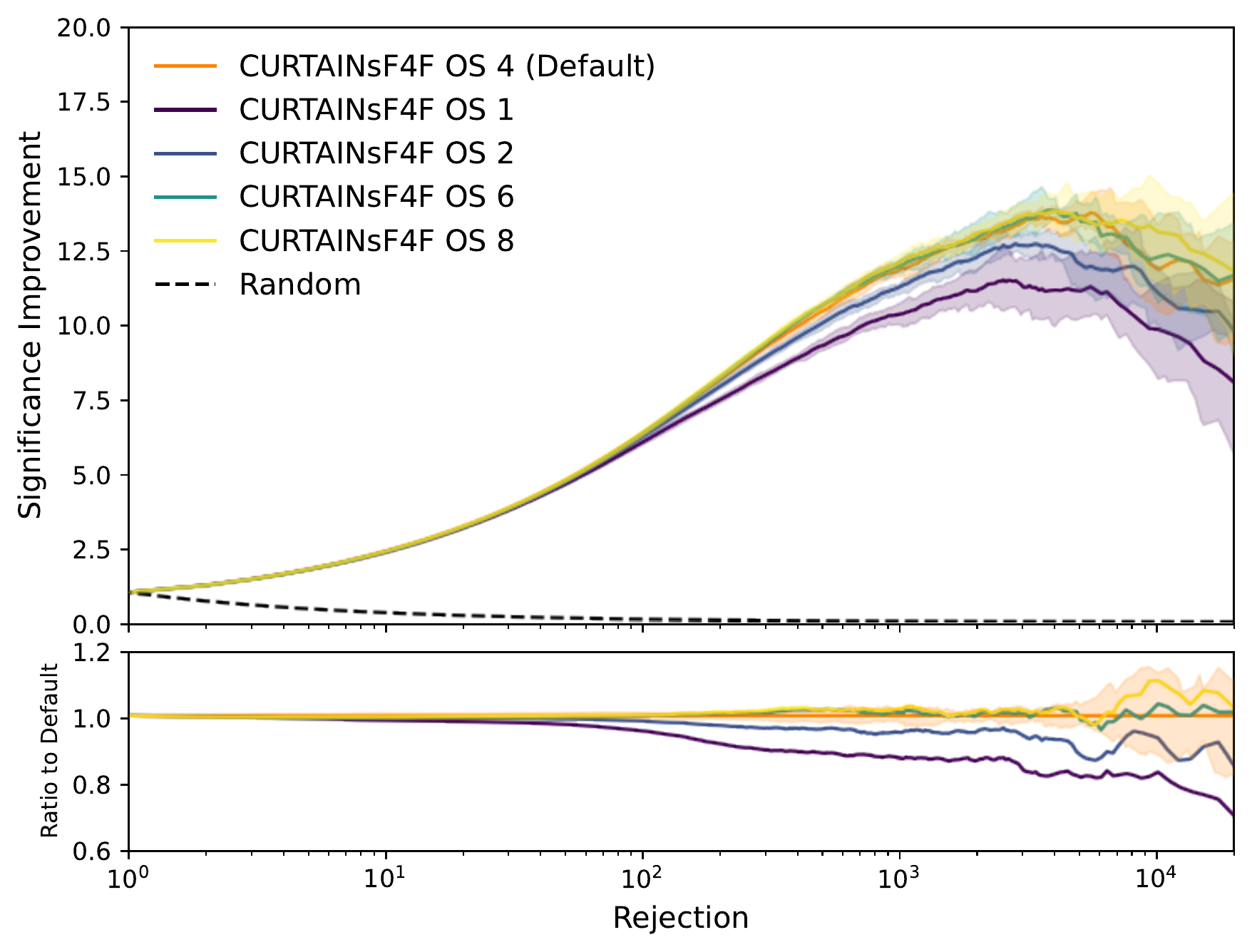}
    \end{subfigure}
    \caption{Background rejection as a function of signal efficiency (left) and significance improvement as a function of background rejection (right) for \FfF trained with varying amounts of oversampling using 200~GeV side-bands
        All classifiers are trained on the sample with 3,000 injected signal events,
        using a signal region $3300\leq\mjj<3700$~GeV.
        The lines show the mean value of fifty classifier trainings with different random seeds with the shaded band covering 68\% uncertainty.
    }
    \label{fig:app:f4foversample}
\end{figure}

An investigation on the sensitivity of \FfF to the amount of oversampling is shown in \cref{fig:app:f4foversample}.
At a factor of four (default) the performance saturates.

In \cref{tab:app:fasttrainingtime} the extrapolated times are computed using the faster top flow but with a new base flow for each signal region.
Although there is a significant time improvement over the default configuration, the efficient implementation still almost three times faster for ten signal regions, and a factor of seven more signal regions can be trained in just over 1000 minutes.

\begin{table}[htbp]
    \caption{The required time to train the base distribution and top flow in \FfF using the faster top flow but a base distribution for each signal region.
        The base distribution and top flow trained on 200~GeV side-bands.
        % The efficient configuration has a single base distribution trained on all data, and a top flow trained on 200~GeV side-bands and optimised for the fastest training time.
        The models are trained on the same hardware with epoch and total training time representative of using an \nvidia~RTX~3080 graphics card.
        An extrapolation of the required total time to train a complete \FfF model for one and ten signal regions are also shown for the two configurations.
        $^\dagger$Timing is for the nominal side-bands, this would vary as the signal region changes due to total number of training events.}
    \label{tab:app:fasttrainingtime}
    \centering
    \begin{tabularx}{0.8\textwidth}{R  Y Y  Y}
        \toprule
        & Time / epoch [s] & $N$ epochs & Total time [min] \\
        \midrule
        \multicolumn{4}{c}{Faster}                                                                                     \\
        \midrule
        Base                                                           & 32.4$^\dagger$   & 100        & 54               \\
        % \midrule
        Top flow                                                       & 21.3$^\dagger$   & 20         & 7                \\
        \midrule
        \multicolumn{3}{r}{One Signal Region}                          & 61                                              \\
        \multicolumn{3}{r}{(Extrapolated$^\dagger$) Ten Signal Region} & 610                                              \\
        \multicolumn{3}{r}{(Extrapolated$^\dagger$) 17 Signal Region}  & 1037                                             \\
        \bottomrule
    \end{tabularx}
\end{table}

\clearpage
\section{Hyperparameters}

% \begin{table}[htbp]
%     \caption{Hyperparameters for training the flows in \FfF.}
%     \begin{tabular}{r c c c}
%         \toprule
%         % & \multicolumn{3}{c}{Network} \\% \multicolumn{4}{c}{Flow}\\
%         % \midrule
%         & Base distribution & Top flow (default) & Top flow (efficient)\\
%         \midrule
%         Number of RQ splines & 4 & 4 & 2\\
%         Number of bins per spline & 4 & 4 & 6\\
%         MADE blocks per spline & 2 & 2 & 6\\
%         Hidden nodes per MADE block & 32 & 32 & 64\\
%         % \midrule
%         % Training \\%\multicolumn{4}{c}{Training}\\
%         \midrule
%         Number of epochs & 100 & 100 & 20\\
%         Batch size & 256 & 256 & 256\\
%         Optimiser & Adam & Adam & Adam\\
%         Initial learning rate & 1e-4 & 1e-4 & 1e-4\\
%         Cosine annealing & True & True & True\\
%         \bottomrule
%     \end{tabular}
% \end{table}

\begin{table}[htbp]
    \caption{Hyperparameters for training the flows in \FfF.}
    \begin{tabular}{r c c c}
        \toprule
        % & \multicolumn{3}{c}{Network} \\% \multicolumn{4}{c}{Flow}\\
        % \midrule
                                  & Base distribution & Top flow (default) & Top flow (efficient) \\
        \midrule
        Number of RQ splines      & 10                & 8                  & 2                    \\
        Number of bins per spline & 4                 & 4                  & 6                    \\
        Transformation            & Autoregressive    & Coupling           & Coupling             \\
        Blocks per spline         & 2                 & 2                  & 6                    \\
        Hidden nodes per block    & 128               & 32                 & 64                   \\
        % \midrule
        % Training \\%\multicolumn{4}{c}{Training}\\
        \midrule
        Number of epochs          & 100               & 100                & 20                   \\
        Batch size                & 256               & 256                & 256                  \\
        Optimiser                 & Adam              & Adam               & Adam                 \\
        Initial learning rate     & 1e-4              & 1e-4               & 1e-4                 \\
        Cosine annealing          & True              & True               & True                 \\
        \bottomrule
    \end{tabular}
\end{table}

\nolinenumbers

\end{document}